\documentstyle[12pt,epsfig]{article}
\begin{document}

\begin{titlepage}
\rightline{March 2004}
\rightline{astro-ph/0403043}
\vskip 3cm
\centerline{\large \bf 
Exploring the mirror matter interpretation of the }
\vskip 0.3cm
\centerline{\large \bf 
DAMA experiment: Has the dark matter problem been solved? }

\vskip 2.2cm
\centerline{R. Foot\footnote{
E-mail address: foot@physics.unimelb.edu.au}}

\vskip 0.7cm
\centerline{\it School of Physics,}
\centerline{\it University of Melbourne,}
\centerline{\it Victoria 3010 Australia}
\vskip 2cm
\noindent
The self consistency between the impressive DAMA annual modulation
signal and the differential energy spectrum is
an important test for dark matter candidates.
Mirror matter-type dark matter passes this test while other
dark matter candidates, including standard (spin-independent)
WIMPs and mini-electric
charged particle dark matter, do not do so well.  
We argue that the unique
properties of mirror matter-type dark matter seem to be
just those required to fully explain the data, suggesting
that the dark matter problem has finally been solved.

\end{titlepage}


Recently it was pointed out\cite{footdama} that the 
impressive DAMA/NaI 
annual modulation signal\cite{dama,dama2} can be simply explained if
dark matter is identified with mirror matter. The purpose
of this article is to explore further this explanation of the
DAMA/NaI experiment. 

Recall, mirror matter is predicted to exist if nature exhibits
an exact unbroken mirror symmetry\cite{flv} (for
reviews and more complete set of references, see Ref.\cite{review}). 
For each type of ordinary
particle (electron, quark, photon etc) there is a mirror partner
(mirror electron, mirror quark, mirror photon etc), 
of the same mass. The two sets of particles form 
parallel sectors each with gauge symmetry $G$
(where $G = SU(3) \otimes SU(2) \otimes U(1)$ in the 
simplest case)
so that the full gauge group is $G \otimes G$.
The unbroken mirror symmetry maps
$x \to -x$ as well as ordinary particles into mirror
particles. Exact unbroken time reversal symmetry
also exists, with standard CPT identified as the product
of exact T and exact P\cite{flv}.

Mirror matter is a rather obvious candidate 
for the non-baryonic dark matter in the Universe because:
\begin{itemize}
\item
It is well motivated from fundamental physics
since it is required to exist if parity and time reversal
symmetries are exact, unbroken symmetries of nature.
\item
It is necessarily dark and stable. Mirror baryons have
the same lifetime as ordinary baryons and couple to mirror
photons instead of ordinary photons.
\item
Mirror matter can provide a suitable framework for
which to understand the large scale structure of the 
Universe\cite{comelli}.
\item
Recent observations from WMAP\cite{wmap} and other experiments suggest
that the cosmic abundance of non-baryonic dark matter is
of the same order of magnitude as ordinary matter $\Omega_{dark} 
\sim \Omega_{b}$. A result which can naturally occur if
dark matter is identified with mirror matter\cite{fvwmap}.
\end{itemize}

Ordinary and mirror particles interact with each
other by gravity and via the photon-mirror
photon kinetic mixing interaction:
\begin{eqnarray}
{\cal{L}} = {\epsilon \over 2} F^{\mu \nu} F'_{\mu \nu} 
\label{km}
\end{eqnarray}
where $F^{\mu \nu}$
($F'_{\mu \nu}$) is the field strength tensor for electromagnetism
(mirror electromagnetism)
\footnote{Given the 
constraints of gauge invariance, renomalizability and mirror
symmetry it turns out\cite{flv} 
that the only allowed non-gravitational interactions
connecting the ordinary particles with the mirror particles
are via photon-mirror photon kinetic mixing 
and via a Higgs-mirror Higgs quartic
coupling, ${\cal{L}} = \lambda \phi^{\dagger} \phi \phi'^{\dagger}
\phi'$. If neutrinos have mass, then ordinary - mirror
neutrino oscillations may also occur\cite{flv2}.}. 
Photon-mirror photon mixing causes 
mirror charged particles to couple to 
ordinary photons with a small effective
electric charge, $\epsilon e$\cite{flv,hol,sasha}.

Interestingly, the existence
of photon-mirror photon kinetic mixing
allows mirror matter-type dark matter
to explain a number of puzzling
observations, including the Pioneer spacecraft anomaly\cite{p1,p2},
anomalous meteorite events\cite{fy,doc}
and the unexpectedly low number of small craters on the
asteroid 433 Eros\cite{fm,eros}. It turns out that these 
explanations and other experimental constraints\cite{fg,ortho,bader} 
suggest that
$\epsilon$ is in the range 
\begin{eqnarray}
10^{-9} \stackrel{<}{\sim} |\epsilon | \stackrel{<}{\sim} 
5\times 10^{-7}.
\label{range}
\end{eqnarray}
However, the strongest direct experimental evidence for mirror
matter-type 
dark matter comes from the impressive DAMA/NaI
experiment\cite{dama,dama2}.
This experiment, located $1.5$ km underground
in the Gran Sasso Laboratory, is designed 
to directly detect dark matter
particles using a NaI target.
This experiment has been in operation for about a decade and
the DAMA collaboration have
claimed discovery of dark matter.
The approach of the DAMA/NaI experiment is to measure nuclear recoils
of Na, I atoms due to the interactions of dark matter particles.
Because of the Earth's motion around the sun, the
interaction rate should experience a small annual modulation:
\begin{eqnarray}
A\cos 2\pi (t - t_0)/T.
\end{eqnarray}
According to the DAMA analysis\cite{dama}, they indeed find
such a modulation over 7 annual cycles at more than 6$\sigma$
C.L. (in the 2-6 keVee energy range):
\begin{eqnarray}
A = 0.019 \pm 0.003 \ {\rm cpd/kg/keV}.
\label{a}
\end{eqnarray}
Furthermore, their data is beautifully self consistent,
because their data
fit gives $T = 1.00 \pm 0.01$ years and
$t_0 = 144 \pm 22$ days, consistent with the expected values.
[The expected value for $t_0$ is 152 days (2 June), where the
Earth's velocity reaches a maximum with respect to the galaxy].
This self consistency is highly non-trivial: there is
simply no reason why their data should contain a periodic
modulation or why it should peak near June 2. In fact, the
estimated systematic errors are several orders 
of magnitude too 
small to account for the signal\cite{dama2,eur}. It seems that DAMA
have indeed discovered dark matter as they have claimed.

It was pointed out in Ref.\cite{footdama} that the
DAMA experiment is sensitive to mirror matter-type dark matter.
Halo mirror atoms can elastically scatter off ordinary atoms via
the photon-mirror photon kinetic mixing interaction, Eq.(\ref{km}).
The DAMA experiment is not particularly sensitive to very
light dark matter particles such as mirror hydrogen and mirror
helium. Impacts of these atoms (typically) do not
transfer enough energy to give a signal above the detection
threshold\cite{footdama}. The next most abundant element is expected to
be mirror oxygen (and nearby elements). 
A small mirror iron component is also possible.
Interpreting the DAMA annual modulation signal in terms
of $O', Fe'$, we found that\cite{footdama}:
\begin{eqnarray}
|\epsilon | \sqrt{ {\xi_{O'} \over 0.10} +
{\xi_{Fe'} \over 0.026}} 
\simeq 4.8^{+1.0}_{-1.3} \times 10^{-9}
\label{dama55}
\end{eqnarray}
where the errors denote a 3 sigma allowed range and 
$\xi_{A'} \equiv \rho_{A'}/(0.3 \ {\rm GeV/cm^3})$ 
is the $A'$ proportion (by mass) of the halo dark matter.

The value of $\epsilon$ suggested by the DAMA experiment, 
Eq.(\ref{dama55}),
would also have implications for the orthopositronium 
system\cite{fg}. The current experimental situation,
summarized in Ref.\cite{ortho,bader}, implies that
$|\epsilon | \stackrel{<}{\sim} 5\times 10^{-7}$,
which is easily consistent with, Eq.(\ref{dama55}).
Importantly, a new orthopositronium experiment
has been proposed\cite{bader} which can potentially
cover much of the $\epsilon$ parameter space suggested by the
DAMA experiment. Such an experiment is
very important -- not just as a check of the
mirror matter explanation -- but also because dark matter experiments are
sensitive to $\epsilon \sqrt{\xi_{A'}}$ and an independent
measurement of $\epsilon$ would allow the extraction of
$\xi_{A'}$ values.

In Ref.\cite{footdama} we found that a DAMA/NaI
annual modulation signal dominated by the $Fe'$ component,
is experimentally disfavoured for three independent reasons:
a) it predicts a differential energy spectrum rate
larger than the measured DAMA/NaI rate. 
b) potentially leads to a significant diurnal effect c) should
have been observed in the CDMS experiment.
Thus it seems probable that the
mirror oxygen component dominates the 
DAMA annual modulation signal, which from Eq.(\ref{dama55})
means that $\xi_{O'} \stackrel{>}{\sim} 4\xi_{Fe'}$.
In this case there are no significant problems.

If the DAMA signal is dominated by $O'$, then things
depend on only one parameter, $\epsilon \sqrt{\xi_{O'}}$.
This parameter is fixed from the annual modulation
signal, Eq.(\ref{dama55}), giving a definite prediction for the
differential energy spectrum in the DAMA experiment [up to the
15\% 1-sigma experimental uncertainty in the annual
modulation amplitude, $A$, Eq.(\ref{a})].
The prediction is shown in {\bf Figure 1}, along
with the DAMA/NaI data (obtained from figure 1 of
Ref.\cite{dama5})\footnote{
See Ref.\cite{footdama} for details of the cross section, form
factors and rate equations used (including effects
of energy resolution, quenching factors etc).}.
Note that for the
annual modulation analysis, the DAMA collaboration
have cautiously adopted a 2 keV software energy threshold.
Nevertheless, for the differential energy spectrum, they have
published data going down to 0.8 keVee\cite{dama5}. 
In general I support
caution, however on this occasion we will
be adventurous; we will use the published data without
further reservations, with the aim of pushing existing
information to the limit.

\vskip 0.3cm
\centerline{\epsfig{file=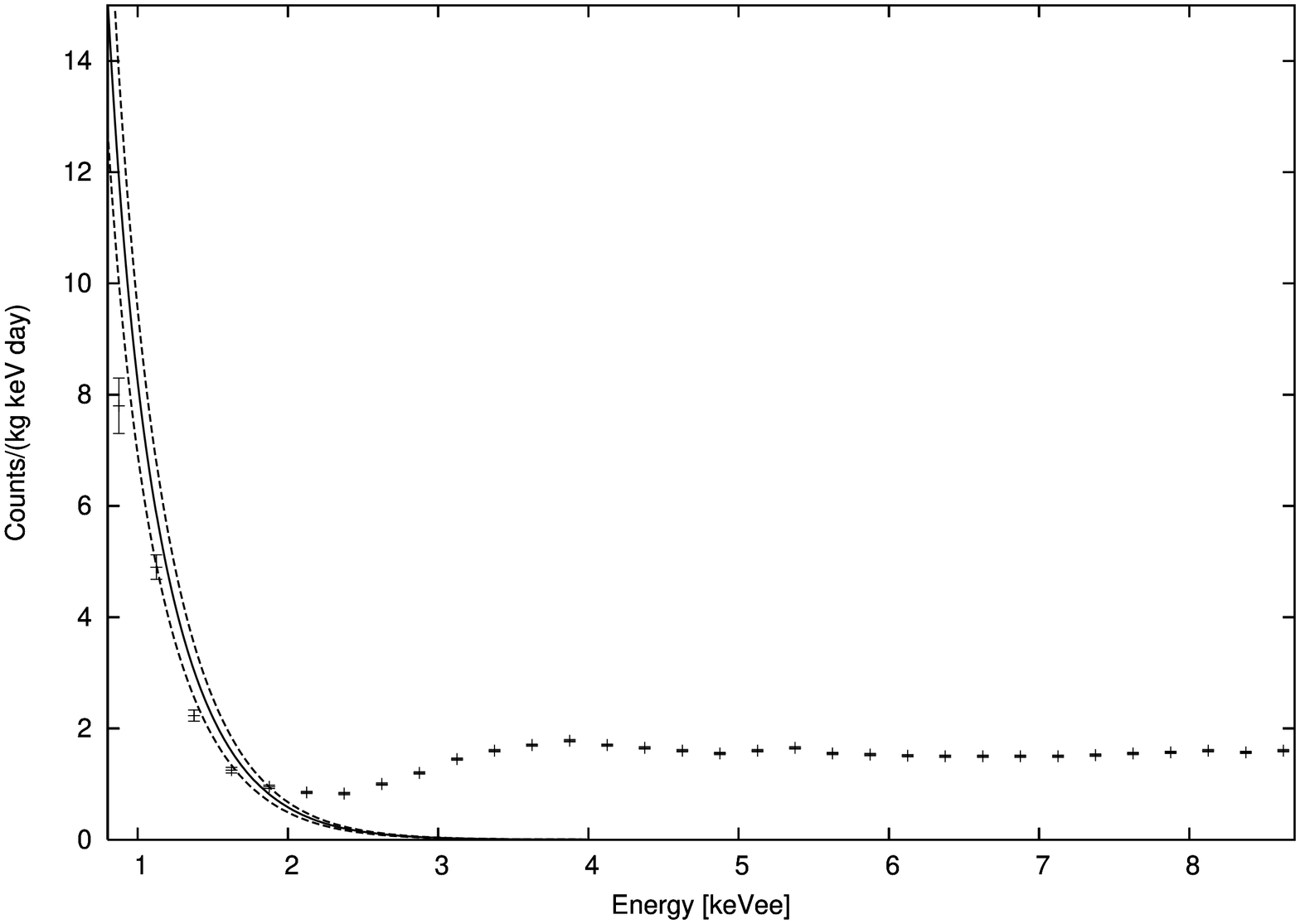,angle=270,width=14.0cm}}
\vskip 0.2cm
{\small \noindent Figure 1: DAMA/NaI differential
energy spectrum prediction for $He'$ dominated
halo with a small $O'$ subcomponent (with
$v_0 (He') = 2v_0 (O') = {220 \over \sqrt{3}}$ km/s). The value of
$\epsilon \sqrt{\xi_{O'}}$ is fixed by the annual
modulation signal, with $A = 0.019$ cpd/kg/keV (central
value) solid line,
$A = 0.022$ cpd/kg/keV (1-sigma upper limit) upper dashed line, 
$A = 0.016$ cpd/kg/keV (1-sigma lower limit) bottom
dashed line.
Also shown is the DAMA/NaI data\cite{dama5}.}
\vskip 1.0cm
\noindent
The fit of the prediction with the experiment is striking
for low recoil energies, as shown in {\bf Figure 2}. 
\vskip -0.2cm
\centerline{\epsfig{file=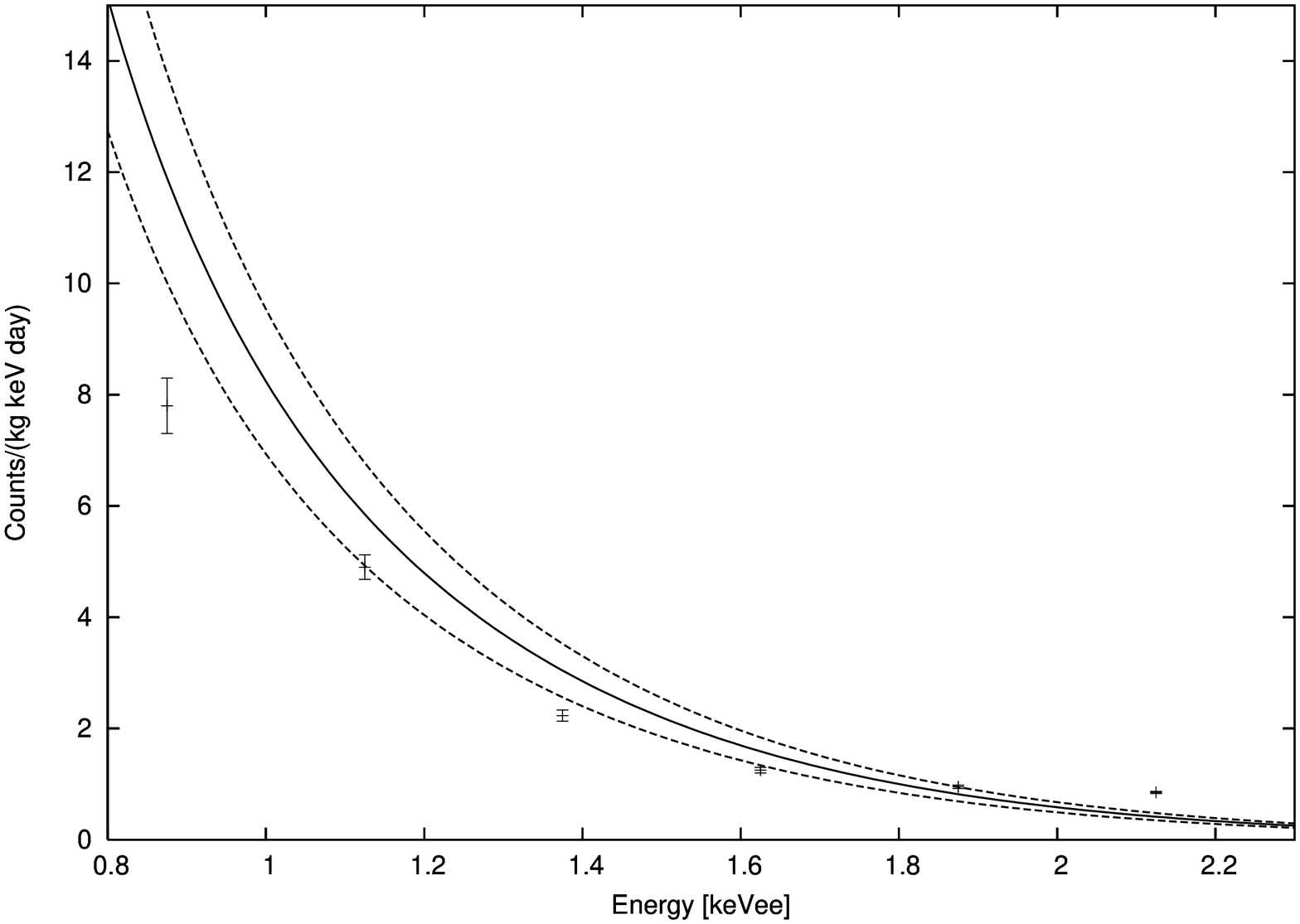,angle=270,width=15cm}}
\vskip 0.4cm
{\small \noindent
Figure 2: Same as Figure 1, but low recoil energy region magnified.}

\vskip 1.0cm
\noindent
It suggests a background
contribution which is roughly constant above 3 keV
and fades to zero below 2 keV, which seems plausible
given the dip in the data at $\sim 2$ keV. 

Our analysis has assumed that the halo has the standard isothermal
Maxwellian distribution (see the appendix for 
details): 
\begin{eqnarray}
f_{A'}(v)/k = (\pi v_0^2)^{-3/2} \ exp[-v^2/v_0^2] .
\end{eqnarray}
In the case of a halo with mass dominated by one
component, which we have assumed is $He'$ (which
is suggested by mirror BBN arguments\cite{comelli}), then
\footnote{In the earlier paper, Ref.\cite{footdama},
we set $v_0 (He') \simeq v_{rot} \approx 220$ km/s.
However, this neglects the pressure due to the mirror electrons,
which are important since $He'$ should be fully
ionized in the halo, given that $T \sim $ keV (see
the appendix for details).}
\begin{eqnarray}
v_0 (He') \simeq {v_{rot} \over \sqrt{3}} \approx {220 \over \sqrt{3}} \ {\rm km/s}.
\end{eqnarray}
Even if the mass of the halo is assumed to be dominated by $He'$, 
there are, of course, still important
astrophysical uncertainties here.
We consider a range for $v_0 (He')$ of (c.f. Ref.\cite{koch}):
\begin{eqnarray}
170 \ {\rm km/s} \stackrel{<}{\sim} \sqrt{3}v_0 (He') \stackrel{<}{\sim} 270\ 
{\rm km/s}.
\label{range2}
\end{eqnarray}
Since the mirror ions/electrons are in
thermal equilibrium with each other,
knowledge of $v_0$ for $He'$ will fix $v_0$ for the 
other elements: 
\begin{eqnarray}
v_0 (A') = v_0 (He') \sqrt{M_{He'}/M_{A'}}\ .
\end{eqnarray}

We now study the effect of varying $v_0$ 
on the mirror matter DAMA prediction, first by retaining
the assumption of $He'$ dominant mass halo and considering
the variation of $v_0 (He')$ given in Eq.(\ref{range2}).
In this case, 
$\sqrt{3}v_0 (He') = 220 \pm 50$  km/s implies 
$\sqrt{3}v_0 (O') = 110 \pm 25$ km/s.  
In going from $\sqrt{3}v_0 (He') = 170$ km/s to $\sqrt{3}v_0 (He') = 270$ km/s
we find that the central value for $\epsilon \sqrt{\xi_{O'}}$
from the annual modulation signal
changes by only $\sim 10 \%$.
In {\bf Figure 3} we examine the effect of varying $v_0$
on the predicted differential energy spectrum 
(while keeping $\epsilon \sqrt{\xi_{O'}}$
fixed by the annual modulation signal, $A = 0.016$ cpd/kg/keV).
As the figure shows, the effect is very slight, with
negligible modification to the predicted rate.

\vskip -0.2cm
\centerline{\epsfig{file=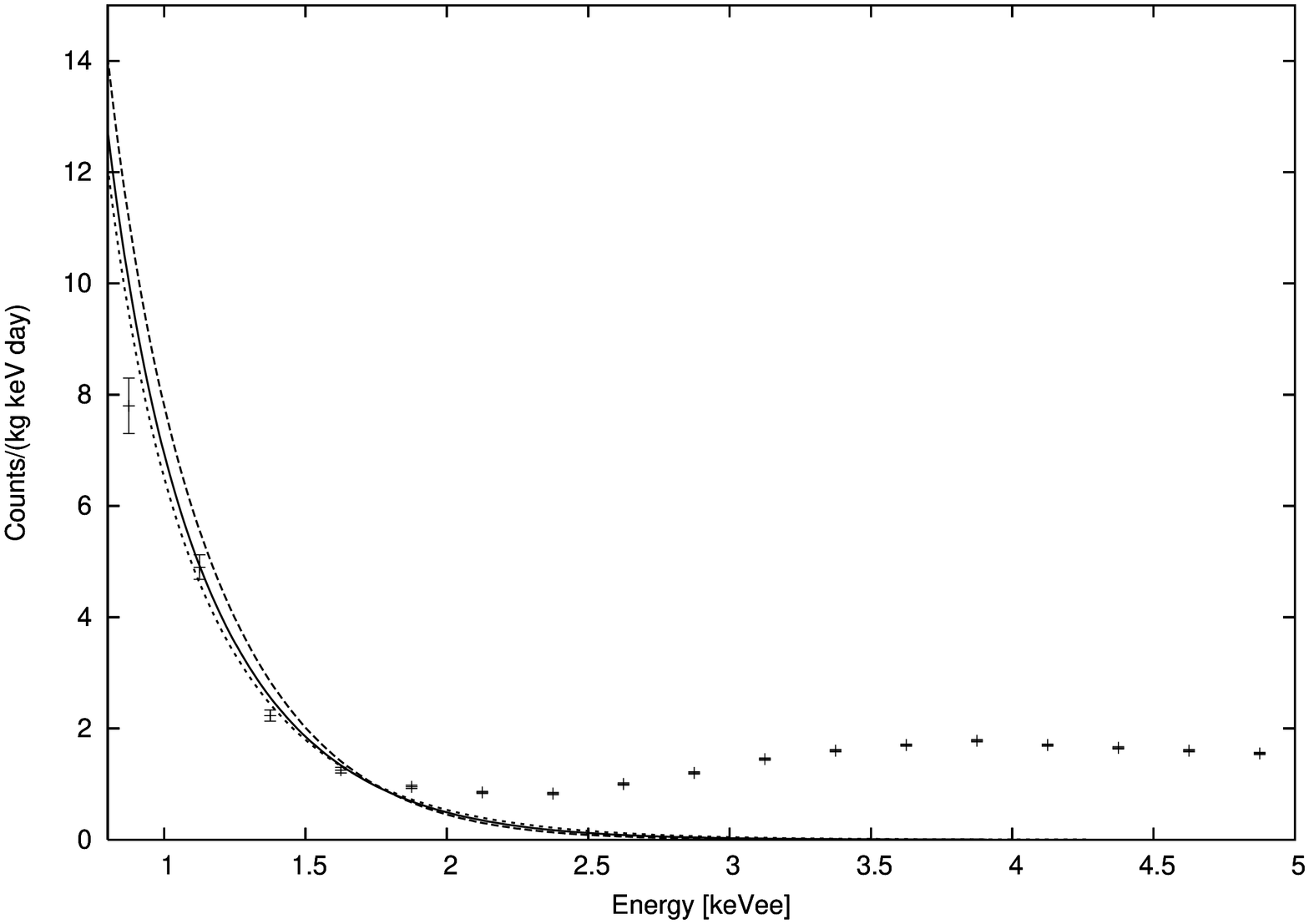,angle=270,width=15cm}}
\vskip 0.4cm
{\small \noindent
Figure 3: 
DAMA/NaI differential energy spectrum prediction, assuming that the
mass of the halo is dominated by $He'$ 
with a small $O'$ subcomponent. The value of
$\epsilon \sqrt{\xi_{O'}}$ is fixed by the DAMA annual
modulation signal (taking the 1-sigma lower value, 
$A = 0.016$ cpd/kg/keV), 
with $\sqrt{3}v_0 (He') = 220$ km/s (solid line),
$\sqrt{3}v_0 (He') = 170$ km/s (dashed line) and
$\sqrt{3}v_0 (He') = 270$ km/s (dotted line).}

\vskip 1.0cm
\noindent
A more extreme variation in $v_0 (O')$ is possible if we vary
the dominant halo component. In particular,
consider the case where the
mass of the halo is dominated by $H'$ rather
than $He'$. This would mean that $\sqrt{2}v_0 (H') \simeq v_{rot} \approx 220$ km/s 
(see appendix) and 
\begin{eqnarray}
v_0 (O') = v_0 (H') \sqrt{M_{H'}/M_{O'}}
\approx {55 \over \sqrt{2}}\ {\rm km/s}. 
\end{eqnarray}
This case, along with the extreme case of a pure $O'$ halo,
with $v_0 (O') = 220$ km/s (which we give just for curiousity)
is shown in {\bf Figure 4}.
\vskip 0.3cm
\centerline{\epsfig{file=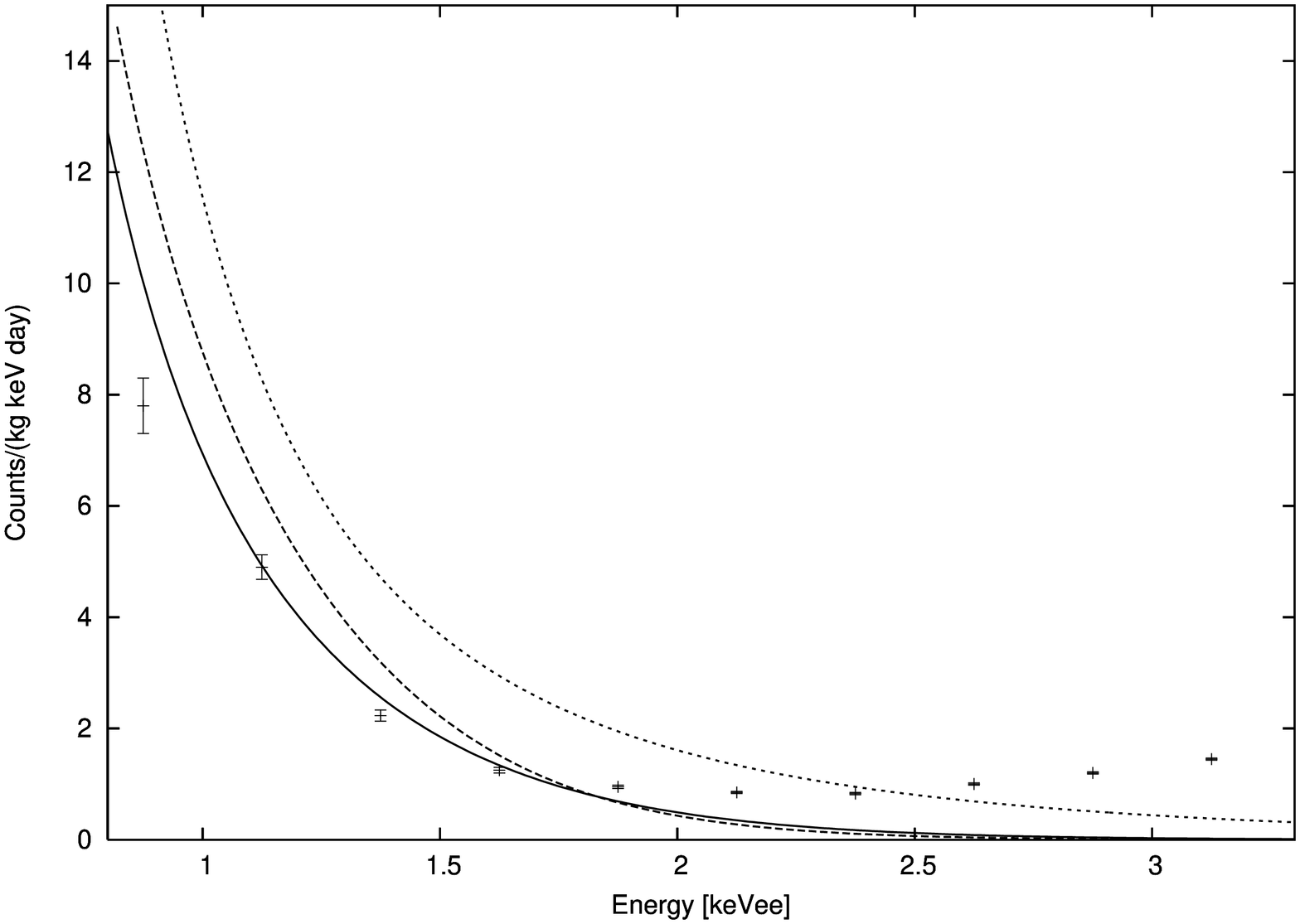,angle=270,width=15cm}}
\vskip 0.3cm
{\small \noindent
Figure 4: Examing the effect of varying the dominate
halo component. In each case we assume that the
halo contains a mirror oxygen subcomponent
but the mass of the halo is dominated by
a) mirror helium (solid line) 
b) mirror hydrogen (dashed line)
and c) containing only mirror oxygen (dotted line, top).
In each case the value of $\epsilon \sqrt{\xi_{O'}}$ is
fixed by the annual modulation signal ($A = 0.016$ cpd/kg/keV). }
\vskip 1.0cm
\noindent
Although the data seems to prefer a dominant $He'$ halo,
the case of a dominant $H'$ halo is certainly still possible, after
taking into account possible uncertainties.
However, a pure $O'$ halo appears to be completely ruled out.
It is of course, theoretically highly unlikely too, but it
does illustrate that the nice fit of the $He'$ case (and to
some extent, the $H'$ case) with small $O'$ subcomponent is
quite non-trivial.  It relies 
on the effect of 
the dominant light elements ($He'$ and/or $H'$ and the
mirror electrons)
to make $v_0(O') \ll 220$ km/s. 

So far we have examined mirror matter dark matter because
it seems to have the right properties to explain
the experimental data. It also seems to be theoretically favoured,
since it arises from just one fundamental hypothesis: that is the
particle interactions respect
an exact, unbroken mirror symmetry.
We now briefly examine
alternative, but less elegant (from a particle physics
point of view!) dark matter candidates. In particular
imagine that the non-baryonic dark matter is composed of
mini-charged particles (and antiparticles)
of electric charge $\pm \epsilon e$ and
mass $M_X$. This case may look superficially similar 
to the mirror matter case, but with
more freedom since the mass, $M_X$, can be arbitrary. 
However, since in this case the halo would have only one
mass component we would expect $v_0 (X) \approx 220$ km/s.
Again we can fit the DAMA annual
modulation signal, but such a model is not consistent
with the measured differential energy spectrum. We illustrate this in
{\bf Figure 5}. 
This clearly demonstrates the necessity of a multi-component halo
to explain the data -- which naturally occurs in the mirror
matter model.
\vskip 0.01cm
\centerline{\epsfig{file=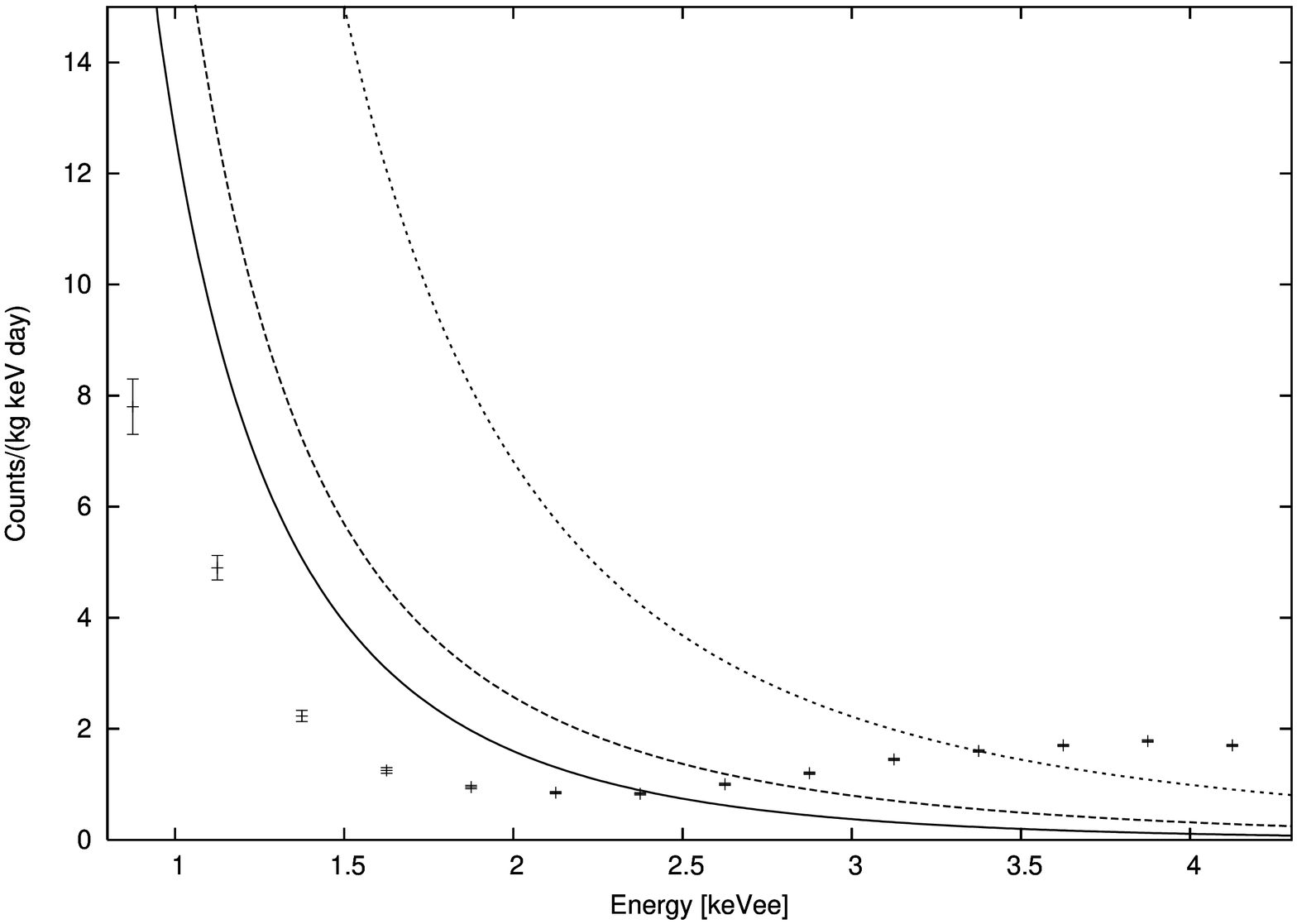,angle=270,width=15cm}}
\vskip 0.3cm
{\small \noindent
Figure 5: DAMA/NaI
differential energy spectrum
prediction for mini-electric charged particle dark
matter (assuming standard halo with $v_0 = 220$ km/s), 
corresponding to the three cases a) $M_X = 40$ GeV (dotted line),
b) $M_X = 20$ GeV (dashed line)
and c) $M_X = 12$ GeV (solid line).
The latter value is actually the best fit to the
measured differential energy spectrum [for
$5 < M_X/GeV < 100$].} 
\vskip 0.9cm
\noindent

Finally in {\bf Figure 6} we compare mirror matter dark matter
with standard spin independent (SI) WIMP dark matter.
Recall WIMP dark matter is quite different from
mirror matter dark matter because its cross section
is point-like:
\begin{eqnarray}
{d\sigma \over dE_R} &=& {
2\pi \epsilon^2 \alpha^2 Z^2 Z'^2 
\over M_A E_R^2 v^2}
F^2_{A} (E_R)
F^2_{A'} (E_R)
\ \ \ {\rm Mirror \ matter\ case}
\nonumber \\
{d \sigma \over dE_R} &=& {2G_F^2 M_A A^2 g^2 \over \pi v^2}
F^2_{A} (E_R)
\ \ \ {\rm WIMP \ case}
\label{xsec}
\end{eqnarray}
where $M_A$ is the target nucleus mass (with mass
number $A$), $g$ is the dimensionless coupling constant
characterising the WIMP-nucleus interaction, $v$ is the
impact velocity (in lab frame) and the $F's$ are the form factors.

\vskip -0.2cm
\centerline{\epsfig{file=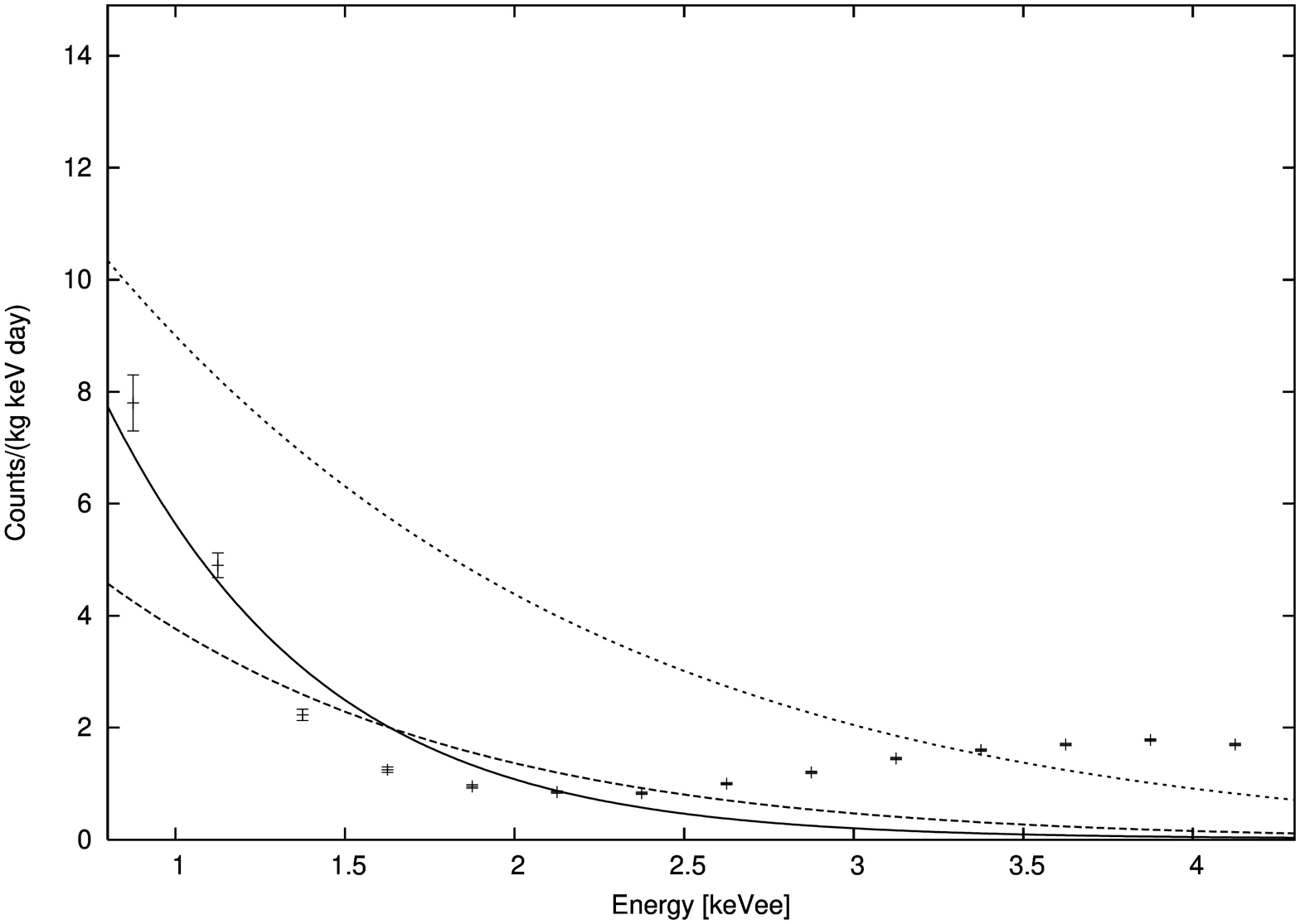,angle=270,width=15cm}}
\vskip 0.4cm
{\small \noindent
Figure 6: 
DAMA/NaI differential energy spectrum prediction assuming standard (SI) WIMP
dark matter.  Three representative cases are considered  
a) 30 GeV WIMP mass (solid line)
b) 50 GeV WIMP mass (dashed line)
and c) 100 GeV WIMP mass (dotted line). 
In each case the cross section is fixed to the value
given by the best fit of the annual modulation signal
($A = 0.019$ cpd/kg/keV).
We have assumed a standard halo with $v_0 = 220$ km/s.
}
\vskip 1.0cm

From Figure 6 we see that after fixing the parameters to the 
DAMA annual
modulation signal the (SI) WIMPs are 
only marginally consistent with
the measured differential energy spectrum and only
if they are light\footnote{
Of course standard (SI) WIMPs are also disfavoured by several
independent dark matter experiments\cite{x,y,z} --
quite {\it unlike} the case of mirror matter-type
dark matter\cite{footdama,ortho}.}.
Note though that we have assumed a standard halo with 
$v_0 = 220$ km/s. Lower values of $v_0$ and
non-standard halo models can improve the situation (see
e.g. \cite{dama2}).
Notice that the predicted differential energy spectrum
is generally flatter than the mirror matter case 
(or mini-charged dark matter case).
Qualitatively, this can be very easily understood from
the $E_R$ dependence of the cross section, Eq.(\ref{xsec}).
[The reason for this different $E_R$ dependence in the
cross section is because of the fundamental interaction:
WIMPs are weakly interacting and mirror matter is electromagnetic].
Evidently mirror matter works for two quite different
reasons a) the cross section is electromagnetic which 
gives the differential energy spectrum the right shape and
b) the value of $v_0 (O')$
is much lower than the anticipated 220 km/s due to the 
effects of the light elements $He'$ and/or $H'$
(which dominates the mass of the halo) and also
the mirror electrons (which contribute to the pressure)
 -- this gives the
spectrum the right normalization.

In conclusion, we have further explored the mirror matter
dark matter interpretation of the DAMA/NaI experiment.
Any proposed explanation of the annual modulation signal
must also be consistent with the measured differential energy
spectrum.
It turns out that this requirement is quite non-trivial,
nevertheless the simplest mirror matter interpretation 
of the DAMA/NaI annual modulation signal --
a $He'$ and/or $H'$ dominated halo with small $O'$ subcomponent --
passes this test.
It is striking that the simplest dark matter candidate coming from
fundamental physics has just the right properties to 
fully explain the DAMA/NaI data. These are strong
reasons to believe that the dark
matter problem has finally been solved. 

\vskip 0.9 cm
\centerline{\large \bf Acknowledgments}
\vskip 0.3cm
\noindent
The author would like to thank D. Cline and
R. Volkas for helpful discussions. This work was
supported by the Australian Research Council.

\vskip 0.9cm

\centerline{\Large \bf Appendix}
\vskip 0.45cm
\centerline{\large \bf Mirror matter-type 
dark matter galactic halo distribution }
\vskip 0.5cm

Assuming a (standard) spherical isothermal dark matter halo, the condition
of hydrostatic equilibrium gives:
\begin{eqnarray}
{dP \over dr} = -\rho g
\label{pr}
\end{eqnarray}
where $P$ is the pressure (for a dilute gas, the pressure
is related to the number density, $n$, via $P = nT$), $\rho$
is the mass density and
$g(r)$ is the local acceleration at a radius $r$.
This acceleration can be simply expressed in terms of the
energy density, via:
\begin{eqnarray}
g = {1 \over r^2}\int^r_{0} G\rho 4\pi r'^2 dr' \ .
\label{fin}
\end{eqnarray}
As is well known, the solution of Eq.(\ref{pr},\ref{fin}) is 
$n \propto 1/r^2$, which gives the required 
flat rotation curve.

The local velocity, $v_{rot} (R_0) \simeq 220$ km/s at our location
in our galaxy is related to the mass density via:
\begin{eqnarray}
v^2_{rot} (R_0) = {G \over R_0}\int^{R_0}_0 \rho 4\pi r^2 dr \ .
\label{vrot}
\end{eqnarray}
If the mass of the galaxy is dominated by the halo component,
then $\rho \propto 1/r^2$. 
The proportionality coefficient can be obtained from
Eq.(\ref{vrot}),
which gives
\begin{eqnarray}
\rho = {v^2_{rot}
\over 4\pi G}{1 \over r^2} \ .
\end{eqnarray}
The gravitational acceleration at a radius, $r$, is then
\begin{eqnarray}
g = {1 \over r}v^2_{rot}
\ .
\end{eqnarray}
Using this equation, we can
solve the hydrostatic equilbrium condition, Eq.(\ref{pr}).
Assuming the simplest case of a
halo made up of a single particle species of mass
$m$, Eq.(\ref{pr}) yields:
\begin{eqnarray}
T = {mv^2_{rot}
\over 2}
\ .
\label{TT}
\end{eqnarray}
Defining the $v_0$ parameter such that the Maxwellian
distribution function [$f(v) = exp\left(-{1\over 2}mv^2/T\right)$]
has the form
\begin{eqnarray}
f(v) = exp\left(-{v^2 \over v_0^2}\right)
\end{eqnarray}
implies that $v_0^2 = 2T/m$. 
Comparing this with Eq.(\ref{TT})
we find that
in the simplest case
of a single component halo, $v_0 = v_{rot}$ (the
standard result).

Repeating this exercise for a spherical isothermal halo composed of 
several components, $H', He', O' ...$ at a common temperature
$T$, we find 
$\rho_{A'} = M_{A'} n_{A'} \propto 1/r^2$, with 
\begin{eqnarray}
T = {\mu M_p  v_{rot}^2 \over 2 }
\end{eqnarray}
and
\begin{eqnarray}
{v_0^2 (A') \over v_{rot}^2} &=&
{\mu M_p \over M_{A'} }.
\label{z5}
\end{eqnarray}
In the above equation,
$\mu M_p$ is the mean mass of the particles comprising the
mirror (gas) component of the halo ($M_p$ is the proton mass).
In determining the mean mass of the particles in the 
halo we must include the 
mirror electron component: at temperatures of order
a keV, $H', He'$ and other light elements are fully ionized.

Let us now consider two illustrative cases: a) the mass
of the gas component of the halo is dominated by $He'$ and
b) the mass of the gas component of the halo is dominated by
$H'$.
In the first case, we have $n_{e'} = 2n_{He'}$ (since $He'$ should
be fully ionized), which means that $\mu M_p \simeq M_{He'}/3$ and from Eq.(\ref{z5})
we have:
\begin{eqnarray}
{v_0^2 (He') \over v_{rot}^2} \simeq {1 \over 3}, \ \ \ [{\rm Case \ a) \ 
Mass \ of \ the \ halo
\ dominated \ by \ He' } ]
\end{eqnarray}
In the second case, we have $n_{e'} = n_{H'}$, which
means that $\mu M_p \simeq M_{H'}/2$ and from Eq.(\ref{z5}):
\begin{eqnarray}
{v_0^2 (H') \over v_{rot}^2} \simeq {1 \over 2}, \ \ \ [{\rm Case \ b) 
\ Mass\ of \
the \ halo
\ dominated \ by \ H' }]
\end{eqnarray}
In each case, the value of $v_0$ for a small $O'$ component
can be obtained from
\begin{eqnarray}
v_0 (O') = v_0 (X') \sqrt{M_{X'}/M_{O'}}\ .
\end{eqnarray}
where $X' = He'$ (first case) and $X' = H'$ (second case).
[Of course the above equation is valid for any $X'$; it
assumes only thermal equilibrium].

\end{document}